\documentstyle[psfig, epsfig, 12pt]{article} 
\input{epsf.sty}

\def\mypagenumber{1}

\def\myend{\end{document}}

\normalsize

\newcounter{sxn}

\newcounter{axn}

\date{}

\newdimen\mybaselineskip
\newcommand{\beeq}{\begin{equation}}
\newcommand{\eneq}{\end{equation}}
\newcommand{\be}{\begin{eqnarray}}
\newcommand{\ee}{\end{eqnarray}}
\newcommand{\bpic}{\begin{picture}}
\newcommand{\epic}{\end{picture}}

\def\d{\partial}
\def\dd{\partial}
\def\la{\raise.16ex\hbox{$\langle$} \, }
\def\ra{\, \raise.16ex\hbox{$\rangle$} }

\def\psibar{ \psi \kern-.65em\raise.6em\hbox{$-$} }
\def\mbar{ m \kern-.78em\raise.4em\hbox{$-$}\lower.4em\hbox{} }

\def\L{ {\cal L} }
\def\ep{\epsilon}

\def\eff{{\rm eff}}

\def\n@space{\nulldelimiterspace=0pt \mathsurround=0pt }
\def\huge#1{{\hbox{$\left#1\vbox to 20.5pt{}\right.\n@space$}}}

\def\myskip{\noalign{\kern 8pt}}
\def\myeqspace{\noalign{\kern 10pt}}

\def\boxit#1{$\vcenter{\hrule\hbox{\vrule\kern3pt
    \vbox{\kern3pt\hbox{#1}\kern3pt}\kern3pt\vrule}\hrule}$}
\def\bigbox#1{$\vcenter{\hrule\hbox{\vrule\kern5pt
     \vbox{\kern5pt\hbox{#1}\kern5pt}\kern5pt\vrule}\hrule}$}

\def\ignore#1{{}}

\begin{document}

\bibliographystyle{unsrt}
\footskip 1.0cm

\thispagestyle{empty}
\setcounter{page}{\mypagenumber}

             

\vspace{2.5cm}
\begin{center}
{\LARGE \bf {Deconfining Phase Transition in 2+1 D:}} \\
\vskip 0.3 cm 
{\LARGE \bf{ the Georgi-Glashow Model. }}\\

\vspace{1.3cm}
{\large Gerald Dunne$^{a,}$\footnote{
dunne@phys.uconn.edu},\hskip 0.2 cm  Ian I.
Kogan$^{b,}$\footnote{kogan@thphys.ox.ac.uk},\hskip 0.2 cm 
 Alex Kovner$^{b,}$\footnote{kovner@thphys.ox.ac.uk},\hskip 0.2 cm
Bayram Tekin$^{b,}$\footnote{tekin@thphys.ox.ac.uk}
}\\
\vspace{.5cm}
$^a${\it Department of Physics, University of Connecticut, Storrs CT
06269, USA}
\\
\vspace{0.5cm} 
$^b${\it Theoretical Physics, University of Oxford, 1 Keble Road, Oxford,
OX1 3NP, UK}\\ 

\end{center}

\vspace*{1.5cm}


\begin{abstract}
\baselineskip=18pt  

\end{abstract}
We analyze the finite temperature deconfining phase transition in
2+1 dimensional Georgi-Glashow model. We show explicitly 
that the transition is due to the restoration of the magnetic $Z_2$
symmetry and that it is
in the Ising universality class.
 We find that
neglecting effects of the charged $W$ bosons leads to incorrect
predictions for the value of the critical temperature and the universality
class of the transition, as well as for various correlation functions
in the high temperature phase. We derive the effective action for the
Polyakov loop in the high temperature phase and calculate the
correlation functions of magnetic vortex operators.

\vfill

PACS: ~ 


 
\newpage



\normalsize
\baselineskip=22pt plus 1pt minus 1pt
\parindent=25pt

\section{Introduction}

The deconfining phase transition in QCD has been a subject of numerous
studies in recent years. Some aspects of the physics
of the high temperature phase appear to be perturbative and can be
studied in a controlled way at asymptotically high
temperatures. However in the phase transition region itself 
the QCD coupling is large and the physics is dominated by the
nonperturbative soft sector. Analytic understanding of this region is
therefore extremely difficult. It thus seems useful to find a simpler
toy model where a similar deconfining phase transition is within the
weak coupling regime and can be studied analytically. Optimistically
one can hope to learn some lessons from such a model which could be of
use in our understanding of real life QCD. 

A model that is very suitable for this purpose is the Georgi-Glashow
model in 2+1 dimensions - Yang Mills theory with a Higgs field
in the adjoint representation.
Over twenty years ago Polyakov \cite{Polyakov} showed that in this theory
the instantons (monopoles) dominate the nonperturbative dynamics
and disorder the Higgs vacuum leading to a 
linear confinement of electric charges even at arbitrarily weak coupling.
This was the first demonstration of the
dual Meissner effect (or dual superconductivity)
in a non-Abelian gauge theory 
where monopoles condense and expel electric fields 
from the vacuum, providing an
example of the enigmatic confinement phenomenon in controllable
circumstances. 
At zero temperature the 3D Yang-Mills-Higgs system is therefore always
in the confining phase, even though classically the confining and
Higgs regimes appear to be separated by a phase transition.

In this paper we will study the theory at finite temperature with
particular emphasis on the  deconfining phase transition.
We note that a similar study appeared in the literature several years
ago \cite{Zarembo}. Our results are both qualitatively and quantitatively  
different from those of
\cite{Zarembo}. In particular we show that the deconfining phase
transition is primarily due to the plasma effects of charged
excitations ($W$ bosons), an effect that was neglected completely in
the analysis of \cite{Zarembo}.
As a result our predictions for both the value of the critical 
temperature and the
universality class of the phase transition are different.

The structure of this paper is the following. In section 2 we review
the model and discuss the two complementary views of the
confinement mechanism: one based on the monopole plasma
\cite{Polyakov} and the other based on vortex condensation 
\cite{'t Hooft, Kovner}. 
In section 3 we discuss the naive expectations for the
deconfinement phase transition which follow from these two approaches.
We show that these expectations are different and explain why
the monopole plasma intuition (the basis of the approach of
\cite{Zarembo}) should be refined in order to discuss the phase
transition properly. In section 4 we describe the renormalization group
analysis of the phase transition. We determine the value of the critical
temperature, the universality class and discuss the crucial role played
by the  electrically charged particles.  
In section 5 we relate our analysis to the standard tool used in the high
temperature phase - the effective potential for the Polyakov line.
We derive this effective potential in the deconfining phase and
show with its help explicitly that the correlation function of vortex
operators at $T>T_c$ decreases exponentially at large distances and
thus that the magnetic $Z_2$ symmetry is restored. Finally in section 6 we
discuss briefly our results.

\section{The model}
We consider the $SU(2)$ gauge theory with a scalar field in the
adjoint representation in 2+1 dimensions.
\be
S= -{1\over 2g^2}\int d^3x \mbox{tr}\left(F_{\mu \nu}F^{\mu 
\nu}\right) 
+ \int d^3x \left[{1\over 2}(D_\mu h^a)^2 +{\lambda
\over  4}(h^a h^a - v^2)^2 \right] 
\label{model1}
\ee
We adopt the notation   $A_\mu = {i\over 2} A^a_\mu \tau^a$, 
$F_{\mu\nu} = \dd_\mu A_\nu  -\dd_\nu A_\mu +[A_\mu, A_\nu]$,
$h = {i\over 2} h^a \tau^a$, and $D_\mu h = \dd_\mu h + [A_\mu, h]$ and
the trace is normalized as $\mbox{tr}(\tau^a \tau^b)= 2 \delta^{ab}$. 

We will be interested throughout this paper in the weakly coupled
regime $v\gg g^2$. In this regime perturbatively the gauge group is
broken to $U(1)$ by the large expectation value of the Higgs field.
The photon associated with the unbroken subgroup is massless whereas
the Higgs and the other two gauge bosons $W^\pm$ are heavy with the masses
\begin{equation}
M^2_H= 2\lambda v^2, \hskip 1.5 cm M^2_W=g^2v^2.
\end{equation}
Thus perturbatively the theory behaves very much like
electrodynamics with spin one charged matter.
The nonperturbative effects however are very important at large
distances. Their effect is that the photon acquires finite mass and
the charged $W^\pm$ become linearly confined at large distances with
nonperturbatively small string tension.

\subsection{The monopole plasma}

The nonperturbative configurations that dominate the generating
functional in the weakly coupled regime are  solutions to the 
classical Euclidean equations of motion,
\be
&D_\mu F^{\mu\nu}  =  [h,D^\nu h] ~~,& \cr
&D_\mu D^\mu h = -\lambda(h^ah^a - v^2) h ~~.&
\label{EulerEq1}
\ee
The perturbative vacuum manifold is $SO(3)/U(1)$ and this has 
a nonzero second homotopy group,
$\Pi_2(SO(3)/U(1))= Z$. The theory thus 
allows topologically non-trivial solutions. 
Polyakov showed \cite{Polyakov} that,
in the vacuum only far separated monopole configurations are
relevant. The unit charge configuration is of the well known 
`t Hooft-Polyakov monopole form,
\be
&&h^a(\vec{x})=\hat x^a h(r) \cr
&&A^a_\mu(\vec{x})= {1\over r} \left[ \epsilon_{a\mu
\nu}\hat{x}^\nu(1-\phi_1)+ \delta^{a\mu}\phi_2 +(r A-\phi_2)\hat{x}^a 
\hat{x}_\mu \right]       
\label{configuration1}       
\ee 
where $\hat x^a = x^a/r$. 
The non-Abelian field strength is
\be
F_{\mu\nu}^a &=&
 {1\over r^2}
  \ep_{\mu\nu b}  \hat x^a \hat x^b (\phi_1^2 + \phi_2^2 -1)
+ {1\over r} (\ep_{a\mu\nu} -  \ep_{\mu\nu b}  \hat x^a \hat x^b )
(\phi_1' + A\phi_2) \cr
\noalign{\kern 10pt}
&& \hskip 4cm + (\delta^{a\nu} \hat x^\mu-\delta^{a\mu} \hat x^\nu) 
{1\over r}(\phi_2' - A\phi_1),
\label{non-abelian field strength}
\ee
where prime denotes differentiation with respect to 3D radial coordinate
$r$. 
The finiteness of the action (in the gauge
$A= \phi_2=0$) requires
$h(0)= 0$ , $h(\infty)= v$, $\phi_1(0)= 1$, 
and  $\phi_1(\infty)= 0$. 

The monopole has a core size of the order of
the vector meson mass. 
Outside the monopole core, the theory is Abelian and 
one can define the Abelian field
strength as,
\be 
F_{\mu\nu} = {h^a\over h} F_{\mu\nu}^a - {1\over h^3} \epsilon_{abc}
             h^a(D_\mu h)^b (D_\nu h)^c.
\ee

The monopoles (instantons) interact via a Coulomb potential in the
Euclidean three dimensional sense, and the partition function of the 
Coulomb gas can be written as a path integral over the ``dual photon
field'' $\phi$. The effective Lagrangian 
for the field
$\phi$ reads
\be
\L_{eff} = {g^2\over32 \pi^2}(\d_{\mu}\phi)^2 + {M^2 g^2\over 16\pi^2}\cos
\phi
\label{sine-Gordon}.
\ee
The dynamically generated photon mass, 
due to the Debye screening of the monopoles, 
is given in terms of the fugacity of the monopoles as
\be
M^2= {16 \pi^2 \xi\over{g^2}}, \hskip 2 cm \xi = \mbox{constant}
{M_W^{7/2}\over g} e^{-{4\pi M_W \over g^2} 
\epsilon({M_H\over M_W})}.    
\ee
$\epsilon({M_H\over M_W})$ is such that  $ 1\leq \epsilon \leq
1.787$ \cite{Prasad}, and $\epsilon(\infty)=1$.
The field $\phi$ is by far the lightest excitation in the theory and
thus eq.(\ref{sine-Gordon}) is the low energy effective Lagrangian of
the theory valid below the energy scale $M_W$.
The fundamental Wilson loop calculated in this effective theory
\cite{Polyakov} has an area law fall off, indicating confinement.

\subsection{The vortex condensation}

A complementary view of confinement in this theory is the spontaneous
breaking of the magnetic $Z_2$ symmetry due to the condensation of
magnetic
vortices\cite{'t Hooft}.

Ignoring monopole effects, the model has the magnetic $U(1)$ symmetry
generated by the Abelian magnetic field. The conserved current is the
dual field strength
\begin{equation}
\tilde F^\mu={1\over 4}\epsilon_{\mu\nu\lambda}F_{\nu\lambda}.
\end{equation}
As discussed in detail in \cite{Kovner}, due to the monopole
contributions this symmetry is anomalous, and only the $Z_2$ subgroup
is conserved. 
The non-anomalous 
$Z_2$ magnetic symmetry group 
is generated \cite{Kovner1} by 
the large spatial Wilson loop which encloses the 
system 
\be
W(C \rightarrow \infty ) = \mbox{exp} {i\over 2}\int d^2 x B(x)
\ee
The order parameter for this $Z_2$ transformation is the operator that
creates an elementary magnetic vortex of flux $2\pi/g$ \cite{'t Hooft,
Kovner}
\be
V(x) = \mbox{exp}{{2\pi i\over g}} \int_C \epsilon_{ij}{h^a\over |h|}
E^a_j(x). 
\ee 
This operator is local, gauge invariant and Lorentz scalar.
Together with the spatial Wilson loop it forms the 
order-disorder algebra 
\be
W(C \rightarrow \infty )V(x) =
 - V(x) W(C \rightarrow\infty)
\ee

It can be shown \cite{Kovner} that the vortex operator condenses in
the vacuum and its expectation value is determined by the gauge
coupling constant $<V>^2=g^2/8\pi^2$.
The low energy theory from this point of view is given by the $Z_2$
invariant theory of the vortex field $V$
\be
{\cal{L}}_{\eff}= \dd_\mu V\, \dd^\mu V^* - 
\lambda ( V\, V^* - {g^2\over 8\pi^2})^2 + 
{M^2\over 4} \{ V^2 + (V^*)^2\}. 
\label{lowlagrangian}
\ee
 The vortex selfcoupling in this effective Lagrangian can be determined as
 \cite{Kovner2} 
\be
\lambda={2\pi^2M_W^2\over g^2}.
\label{couplings2}
\ee

Note that as a low energy Lagrangian, eq.(\ref{lowlagrangian}) is
indeed consistent with eq.(\ref{sine-Gordon}). At weak gauge coupling the
quartic coupling $\lambda$ is very large. In this nonlinear
$\sigma$-model limit the radius of the field $V$ is therefore
frozen to its expectation value. The only relevant
degree of freedom is the phase
\be
V(x) = {g\over {\sqrt{8}\pi}} e^{i\chi}. 
\label{vortex1}
\ee
Substituting this into eq.(\ref{lowlagrangian}) one indeed obtains
precisely eq.(\ref{sine-Gordon}) with the identification $\phi=2\chi$.
It was shown in \cite{Kovner} that this identification is indeed the
correct one, and the factor $2$ follows from the fact that the
magnetic charge
of the monopole is twice the fundamental value relative to the
electric 
charge of the elementary charged excitation
$W$ : $g_Mg_e=4\pi$.

The main difference between eq.(\ref{lowlagrangian}) and
eq.(\ref{sine-Gordon}) is in the description of the charged sector. As
explained in \cite{Kovner}, the charged particles are solitons of the
field $V$ with unit winding number. In the nonlinear $\sigma$-model limit
those are just vortices of the phase $\chi$. Polyakov's effective
Lagrangian eq.(\ref{sine-Gordon}) does not allow such vortex
configurations. In the vortex configuration the field $\chi$ (and
$\phi$) is discontinuous along an infinite cut. The cut contributes to
the kinetic energy term in eq.(\ref{sine-Gordon}), and this
contribution is both ultraviolet and infrared divergent. The
ultraviolet divergence is of no importance by itself, since the theory
is defined with the intrinsic UV cutoff $M_W$. However the infrared
divergence
indisputably puts description of the charged states beyond
eq.(\ref{sine-Gordon}). On the other hand eq.(\ref{lowlagrangian})
recognizes the fact that the field $\chi$ is a phase, and therefore
$2\pi$ discontinuities in it are unphysical and do not cost
energy. Thus vortex configurations in eq.(\ref{lowlagrangian}) are
allowed. Although they are confined dynamically with a linear potential
\cite{Kovner}, 
the
string tension for this potential is a dynamical effect 
proportional to $\xi^{1/2}$ and
has nothing to do with the extraneous cut contribution.

In fact, if anything the Lagrangian eq.(\ref{lowlagrangian})
underestimates the energy of such a vortex. The core energy of the
vortex should be equal to the mass of the charged vector boson
$M_W$, since this is the lightest charged excitation in the theory.
On the other hand the core mass of a vortex in the Lagrangian
eq.(\ref{lowlagrangian}) is given by the UV contribution of the
Coulomb potential. With the cutoff of order $M_W$ this is of order $g^2
\ln M_W/g^2$. This is not surprising, since the mass of the $W$ boson
indeed comes from the distances of order of its Compton wave length,
and is thus well inside the UV cutoff of the effective theory. The
situation can be improved by adding to the effective Lagrangian
a higher derivative term of the Skyrme type
\be
\delta L=\Lambda(\epsilon_{\mu\nu\lambda}\partial_\nu V^*
\partial_\lambda V)^2
\label{skyrme}
\ee
with
\be
\Lambda\propto{\frac{1}{g^4}}{\frac{1}{M_W}}
\ee
With this extra term the core energy of the vortex indeed becomes of
the order of $M_W$.

As we will see in the following, the correct description of the
charged sector is crucial for the understanding of the phase
transition\footnote{We note that this is not the only instance in
which the charged states play important role. For example 
as discussed in \cite{greensite} the presence of these states leads to
the breaking of the string of charge two, which in the Polyakov's
effective Lagrnagian eq.)\ref{sine-Gordon} is strictly stable.}.

\section{Deconfining phase transition - The Rough Guide}

In this section we give a naive discussion of the expected
nature of the deconfining phase transition based on the monopole
versus vortex description of the low energy theory.

\subsection{The monopole binding}

Since at zero temperature the monopole contributions are the only
relevant ones for confinement, one may be tempted to assume that also at
finite, low enough temperature
all other effects are not important. The expected value 
of the transition temperature, as we shall see in a moment is of order
$g^2$, which is much smaller than any mass scale in the theory except
for the photon mass. One could therefore start with the working
hypothesis that the single self-interacting photon field (or equivalently 
the monopole ensemble) should be a valid description of the phase
transition region. 
This is the point of view taken in \cite{Zarembo}, and we now briefly
describe the conclusions it leads to.

The first thing to note is that at finite temperature the interaction
between monopoles is logarithmic at large distances. 
The reason is that the finite temperature path integral is formulated
with periodic boundary conditions in the Euclidean temperature
direction. The field lines are therefore prevented from crossing the
boundary in this direction. The magnetic  
field lines emanating from the monopole have
to bend close to the boundary and go parallel to it. So effectively
the whole magnetic flux is squeezed into two dimensions. Qualitatively the
situation is shown in figure 1.
\begin{center}
\epsfig{file=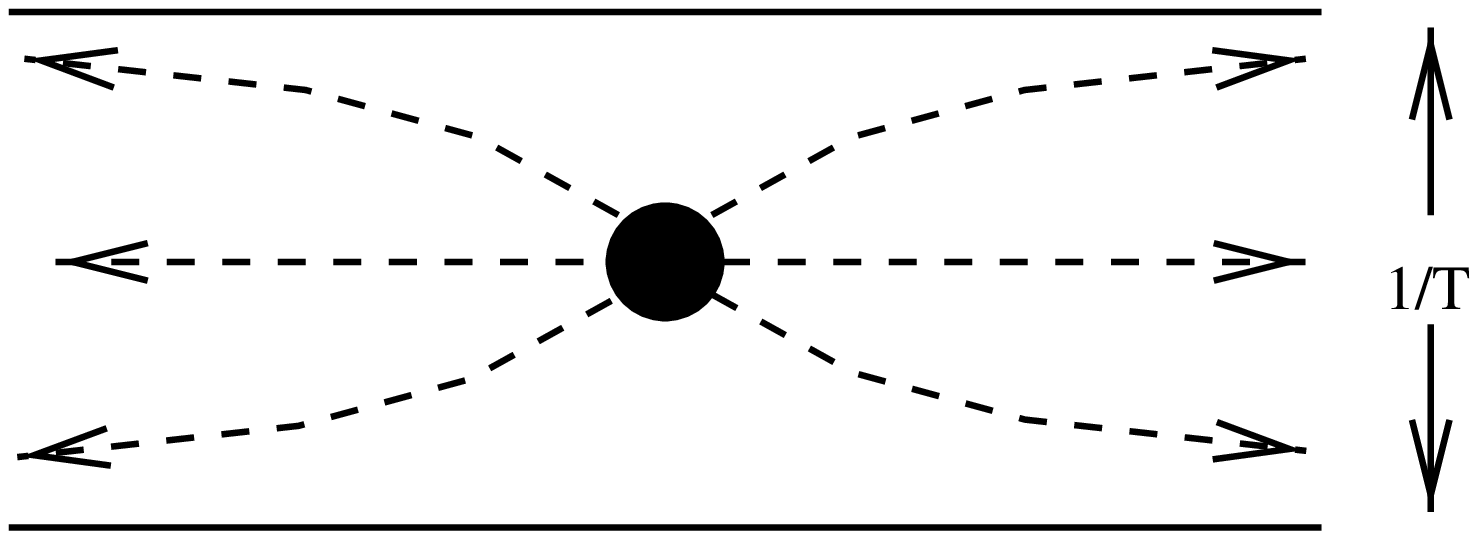, width=10.cm}
\end{center}
\baselineskip=22pt
\centerline{\small{Figure 1: The field of a Monopole-Instanton at finite
temperature. }}
\baselineskip=22pt
The length of the time direction is $\beta=1/T$, and thus clearly the
field profile is two dimensional on distance scales larger than
$\beta$. Two monopoles separated by a distance larger than $\beta$
therefore interact via a two dimensional rather than a three dimensional
Coulomb potential, which is logarithmic at large distances. Since the
density of the monopoles is tiny $\rho_M\propto\xi$, already at
extremely low temperatures $T\propto \xi^{1/2}$ the monopole gas
becomes two dimensional.
The strength of the logarithmic interaction is easily calculated.
The magnetic flux of the monopole should spread evenly in the
compact direction once we are far enough from the monopole core
The field strength should have only components parallel to the spatial
directions.
Since the
total flux of the monopole is $2\pi/g$, the field
strength far from the core is 
$\tilde F_i={T\over g}{x_i\over x^2}$,
and thus the strength of the infrared logarithmic interaction
is $T^2/g^2$.

It is well known \cite{BKT} that the two dimensional Coulomb gas 
undergoes the BKT phase transition. In the usual Coulomb gas, where
the strength of the interaction $\lambda$ does not depend on temperature,
the
particles are bound in pairs at low temperature and unbind at high
temperature $T>T_{BKT}=2\pi\lambda $, where the entropy overcomes the
energy. This is the standard BKT phase transition \cite{BKT} 
which determines the
universality class of $U(1)$ symmetry restoration in 2 dimensions.
In the present case the situation is reversed, since the strength of
the interaction itself depends on the temperature. At low temperature the
interaction is weak, and therefore the particles (monopoles) are
free. As the temperature grows, the interaction becomes stronger until at
$T_{BKT}=2\pi {T^2_{BKT}\over g^2}$ the energy overcomes the
entropy. Above this temperature the monopoles bind into neutral pairs.
Thus
based on this simple picture we expect the theory to undergo a BKT
phase transition at
\be
T_{BKT}={g^2\over 2\pi}
\label{TBKT}
\ee
Below this temperature the photon should be massive, while above this
temperature it should be massless since the cosine term in the
Lagrangian eq.(\ref{sine-Gordon}) is irrelevant.

This is precisely the logic followed in \cite{Zarembo}. One can go
further and perform more quantitative calculations using the
dimensionally reduced Polyakov effective Lagrangian. Dimensional
reduction should be perfectly valid in this theory since the critical
temperature is much larger than the photon mass.
Thus one can consider the two dimensional sine-Gordon theory
\be
\L = {g^2\over32 \pi^2 T}(\d_{i}\phi)^2 + {M^2 g^2\over 16\pi^2 T}\cos
\phi.
\ee
This theory, in full agreement with the previous discussion has a BKT
phase transition at the temperature given by eq.(\ref{TBKT}), above
which the ``dual photon'' field $\phi$ becomes massless.

Moreover, since the 2-d sine-Gordon theory is exactly solvable, 
one can calculate various correlation functions. In particular the
authors of \cite {Zarembo} calculated the string tension, and
found that it exhibits quite strange behavior. At low temperatures it
follows an expected pattern - that is, it decreases with temperature. It
becomes extremely small at temperature much smaller than $T_{BKT}$;
however just before $T_{BKT}$ it starts rising sharply and at $T_{BKT}$
actually diverges.

Although the logic leading to the previous discussion seems quite
sound, there are several puzzles that arise regarding the results.

$\bullet$ The photon becomes massless in the high temperature phase. In
other words, the correlation length in some physical (gauge invariant)
channel becomes infinite at high temperature. This
contradicts our physical intuition, since one expects that at high
temperature any physical system described by a local field theory
should  become maximally disordered with zero correlation length.

$\bullet$ The phase transition is of BKT type, and therefore belongs to
the $U(1)$ universality class. On the other hand we know that the
global symmetry that is restored at the phase transition is the
magnetic $Z_2$ \cite{Kovner1}, and we expect the universality class to
be 2D Ising.

$\bullet$ The divergence of the string tension at the critical point looks
unphysical.

$\bullet$ Finally, as we explained in the previous section, the
effective theory eq.(\ref{sine-Gordon}) does not allow charged
states. It is then difficult to understand in what sense the high
temperature phase can be viewed as deconfined.

\subsection{Magnetic symmetry restoration and the charged plasma}
Let us now take eq. (\ref{lowlagrangian}) as our starting point.
What does one expect from the phase transition in a simple scalar
theory of this type? 
Firstly, clearly one expects the critical temperature to be of order
of the expectation value of the scalar field, and therefore
parametrically
of order $g^2$. At finite
temperature one expects the generation of a positive thermal mass
proportional to the temperature and to the coupling constant. Thus the
thermal contribution to the effective potential should be of the form
\be
\delta_TL=x\lambda TV^*V
\ee
At $T=g^2/4\pi^2x$ the total mass term becomes positive, the VEV
of $V$ vanishes and the phase transition occurs. 
One can  estimate this temperature more precisely using the following
simple
argument. Let us neglect for now the monopole induced term. Then
we are dealing with the $XY$ model at finite temperature.
The dimensional reduction should again be a valid approximation, and
thus essentially we have to analyze a 2D model. Now a 2D $XY$ model
can be mapped into a sine-Gordon theory of a dual field
$\tilde\chi$. Performing this dual transformation\footnote{To fix the
normalization of the kinetic term we should bear in
mind that a vortex corresponds to a
$2\pi$ jump in the field $\chi=2\phi$ rather than the field $\phi$ in
eq.(\ref{sine-Gordon}). See also the discussion following
eq. (\ref{vortex1}).}
,
we find
the Lagrangian
\be
L={T\over 2g^2}(\partial_i\tilde\chi)^2+\mu\cos\tilde\chi
\label{dualsG}
\ee
where $\mu$ is the fugacity of the vortices in the $XY$ model.
This sine-Gordon theory has a phase transition at
\be
T_{XY}={g^2\over 8\pi}.
\ee
Therefore we may expect the magnetic symmetry restoring phase transition
at $T_{XY}$. Although parametrically this temperature is of the same
order as the BKT transition temperature discussed in the previous
subsection, it is four times lower. Another important difference is
that the nature of the phase transition is in fact completely
different.
The phase transition in the model eq. (\ref{dualsG}) is due to the
unbinding of vortices of the field $V$. Above $T_{XY}$ the vortices
are in the plasma phase. These vortices, are as
discussed earlier precisely the charged $W^\pm$ bosons of the original
Georgi-Glashow model. Thus this phase transition is just what one
would naturally call the deconfining phase transition and the vacuum above
$T_{XY}$ is the charged plasma. 

Note that in this discussion we have
neglected completely the effect of monopoles, that is the last term in
the Lagrangian eq. (\ref{lowlagrangian}). Interestingly enough we see
that the plasma phase is reached at a temperature which is much
lower than the $T_{BKT}$ discussed earlier, and thus the monopole
binding is  irrelevant for the dynamics of the
deconfinement.
This is not to say that the presence of the monopoles is 
irrelevant altogether. Clearly,
omitting the monopole induced term we enlarged the symmetry of the
system from $Z_2$ to $U(1)$. Hence the effective $XY$ model
description and the $U(1)$ universality class predicted by
eq.(\ref{dualsG}). The analysis as it stands is correct for noncompact
electrodynamics with charged matter, but not for the Georgi-Glashow model.
Next we remedy this problem.

The effect of the monopoles can be qualitatively understood in a
simple way. Their presence leads to a confining potential between the
charges, that is a linear interaction between the $XY$ model vortices. 
Thus the monopoles suppress the variation of the phase of the field $V$
except inside the confining string, the width of which is given by the
inverse photon mass $d=1/M\propto\xi^{1/2}$. Whether the
presence of the monopoles is important at the ``would be'' 
phase transition point $T_{XY}$
depends crucially on the density of charged particles. If the density
of the charged particles is very high, so that the average distance
between them is smaller than $d$, the presence of the 
monopoles is immaterial since
the phase of $V$ is disordered already on the short distance
scale. However if the density of charges is low, so that the distance
between them is larger than $d$, the presence of the monopoles will
suppress the phase transition. In this case at $T_{XY}$ the vacuum will
not be disordered, but will look like a dilute gas of charged particles
with strings between them. The actual phase transition will then occur
at a higher temperature, where the average distance between the
charges equals the inverse photon mass. 

Interestingly enough in the present
case this temperature turns out to be twice the value of $T_{XY}$ and
therefore still much lower than $T_{BKT}$. We will show this
rigorously in the next section. Qualitatively though this is easy to
understand. The density of charges is proportional to their fugacity
\be
\mu\propto e^{-M_W\over T}
\ee
This should be compared to the square of the photon mass, which is
given by the zero temperature monopole fugacity. In a theory with very
heavy Higgs
\be
\xi\propto e^{-4\pi M_W\over g^2}
\ee
The two are equal (up to subleading corrections) at
\be
T_{GG}={g^2\over 4\pi}
\ee
More generally, when the Higgs is not infinitely heavy, the monopole
fugacity is smaller and thus the transition temperature will be lower
\be
T_c={g^2\over 4\pi\epsilon({M_H\over M_W})}
\ee
For lighter Higgs, $T_c$ gets closer to $T_{XY}$, but is always
greater so that $ T_{XY} < T_c < T_{BKT}$.

In the next section we will present a more complete analysis based
on the renormalization group and exact bosonization. This analysis
confirms the simple picture presented here. Thus all the puzzles
raised in the previous subsection simply disappear.

$\bullet$ The photon never becomes massless.
Even without the monopoles in the plasma phase it acquires the Debye
mass given by the cosine term in eq. (\ref{dualsG}). This mass {\it
rises} with temperature, and thus the physical correlation length
decreases.

$\bullet$ Since the monopole term is still relevant at the phase
transition, the universality class must be $Z_2$. We will see this
explicitly in the next section.

$\bullet$ The analysis of the previous subsection is only valid below
$T_{XY}$, and thus the divergence of the string tension 
at $T_{BKT}>T_{XY}$ has nothing to do with the physics of the Georgi
Glashow model.

$\bullet$ Finally,
the phase transition is driven primarily by the unbinding of charged
particles and thus indeed has all the flavor of a deconfinement
transition.

Let us now turn to a more quantitative analysis of what is stated above.

\section{Deconfining phase transition -  renormalization group
analysis}
 
For a more formal analysis of the phase transition we find it
convenient to use the sine-Gordon formulation in terms of the phase
field
$\chi$
modified to take explicitly into account the finite probability
of the appearance of vortices.
This Lagrangian has the form
\be
\L = { g^2 \over 8 \pi^2 T } (\d_{\mu}\chi)^2 + \zeta\cos 2 \chi 
+ \mu \cos \tilde{\chi}   
\label{sinevortex}
\ee
where $\zeta$ is related to the monopole fugacity $\zeta=\xi/T$ and 
$\tilde \chi$ is the field dual to $\chi$, 
\be
i\d_{\mu}\tilde\chi= {g^2\over 2\pi T} \epsilon_{\mu\nu}\d^\nu \chi.
\label{chid}
\ee
\subsection{The Lagrangian}
The way to derive this Lagrangian is the following. The partition
function of the sine-Gordon model in the presence of one vortex is
\be
Z(x)=\int D[\chi]\exp\{-\int d^2y { g^2 \over 8 \pi^2 T }
(\d_{\mu}\chi-
j_\mu(y,x))^2 + \zeta \cos 2 \chi \}
\label{onev}
\ee
The ``external current'' is
\be
j_\mu(y,x)=2\pi n_\mu(y)\delta(y\in C)
\ee
with $C$ a curve that starts at the location of the vortex (the point
$x$), and goes
to infinity, and $n_\mu$ is the unit normal to this curve.
The insertion of this current forces the derivative of $\chi$ to have
a discontinuity across the curve $C$, so that $\chi$ jumps
by $2\pi$. This forces $\chi$ to have one unit of vorticity concentrated
at the
point $x$.
Note that even though $j_\mu$ explicitly depends on the curve $C$, the
partition function itself does not, since changing the integration
variable
\be
\chi(x)\rightarrow\chi(x)+2\pi , \ \ \ \ \ \ \ x\in S
\ee
where the boundary of $S$ is $C-C'$ is equivalent to changing $C$ into
$C'$ in the definition of the current.
The extra linear term in the exponential in (eq.\ref{onev})
is
\be
i\tilde\chi= { g^2 \over 2 \pi T }\int_C
dx_\mu\epsilon_{\mu\nu}\partial_\nu\chi.
\ee
which is equivalent to (eq.\ref{chid}).
An antivortex at $y$ obviously is created by $-j_\mu$.
To create several vortices one just inserts an external current which
is the sum of the currents which create individual vortices.

A dilute ensemble of vortices and antivortices with (small) fugacity $\mu$
is then given by
\be
Z=\sum_{n,m}{1\over n!}{1\over m!}\mu^{n+m}\int \Pi_idx_i\Pi_jdy_j
Z(x_i,y_j)
\ee
The summation over the number of vortices and antivortices can be
easily performed leading to the partition function with the Lagrangian
eq. (\ref{sinevortex}).
The constant $\mu$ is the vortex fugacity scaled by the effective UV
cutoff imposed on the integration over the coordinates. 
The vortex fugacity of course is none other but the fugacity of the
charged $W$,
\be
\mu=a^{-2} e^{-M_W\over T}.
\ee
The cutoff $a$ is related to the Compton wave length of the $W$ boson,
but a more careful determination of it should take into account the
fact that in the process of dimensional reduction all modes with
frequencies above $T$ have been integrated out. We will not attempt
the determination of $a$, but only note that it is some combination of
the scales $M_W$ and $T\propto g^2$, and as such its value always 
plays a role secondary to the exponential factor of fugacity\footnote{
For the same reason the effective value of the monopole induced mass
$\zeta$ also
changes in the dimensionally reduced theory. This change has been
calculated in \cite{Zarembo}. Again we will not dwell on this finite
``renormalization'' since this effect is subleading in the same sense.}.

An alternative way to derive the
Lagrangian eq. (\ref{sinevortex}) is to start directly from the effective
theory
eq. (\ref{lowlagrangian}) with the extra Skyrme term eq. (\ref{skyrme}).
In the nonlinear $\sigma$-model limit one can cleanly separate the phase
of
the field $V$ into a smooth part $\chi$  and the vortex contribution. The
Skyrme term then is proportional to the energy of one vortex, and with
the Skyrme coupling chosen the way we discussed in Section 2 is just
equal to $M_W$ for a one vortex configuration. The dilute vortex gas
approximation then reproduces again the Lagrangian eq. (\ref{sinevortex}).

\subsection{The renormalization group equations}
Our starting point for the discussion of this section is
eq. (\ref{sinevortex}).
Since both $\xi$ and $\mu$ are small, the importance of different
terms in the Lagrangian is determined by their respective conformal
dimensions calculated in the free theory.
The total conformal dimensions of the operators $:\cos 2 \chi:$ , $:\cos
\tilde{\chi}:$
respectively are, 
\be 
\Delta_\xi=  {4\pi T\over g^2}, \hskip 2 cm   \Delta_\mu=  {g^2 \over 4\pi
T}.
\ee
Thus at low temperature  the charges 
are irrelevant ($\Delta_\mu>2$) and the monopoles are relevant 
($\Delta_\xi<2$) 
and the theory reduces to the sine-Gordon model.
At the temperature $T=g^2/4\pi$ the two operators become equally
relevant since their conformal dimensions are equal
\be
\Delta_\mu=  \Delta_\xi= 1.
\ee

The phase transition point
can be determined from the structure of the fixed points
of the renormalization
group equations. The critical point is the IR unstable fixed point of
the RG flow.
The renormalization group equations of the sine-Gordon 
theory have been studied perturbatively in the 
literature both in the absence \cite{Coleman} 
and in the presence of the vortices \cite{Kadanoff}.
In terms of the dimensionless parameters
\be
t={4\pi\over g^2}T, \ \ \ \ \ \tilde\mu=\mu a^2, \ \ \ \ \ \
\tilde\zeta=\zeta a^2
\ee
the lowest order RG equations read
\be
&&{dt\over d\lambda} = \pi^2(\tilde\mu^2 - t^2 \tilde\zeta^2), \\
&&{d\tilde\mu\over d\lambda} = (2- {1\over t})\,\, \tilde\mu, \\
&&{d\tilde\zeta\over d\lambda} = (2-   t)\,\, \tilde\zeta.
\label{RG}
\ee
The fixed point structure of these equations is simple.

1. The point $T_0$
\be
t=0,\ \ \ \ \tilde\mu=0, \ \ \ \ \tilde\zeta=\infty
\ee
is clearly the zero temperature fixed point. Here the long distance
physics is dominated by the monopole induced mass term.

2. The point $T_\infty$
\be
t=\infty, \ \ \ \ \ \tilde\mu=\infty, \ \ \ \ \ \tilde\zeta=0
\ee
is the high temperature fixed point. Here the infrared properties are
detemined by the charged plasma effects.

3. The point $T_{GG}$
 \be
t=1,\ \ \ \ \ \tilde\mu=\tilde\zeta, \ \ \ \ \tilde\mu=\infty
\ee
This is the IR unstable fixed point. This is precisely the critical point
 that corresponds to the deconfining phase transition. At this point
 both the monopole and the charge plasma induced terms are equally
important.

Note that the fixed line $t>2, \ \ \tilde\zeta=0$ of massless theories
which corresponds to
the ``confined monopole plasma'' is not present, since for $t>2$ the
charged plasma induced mass term is strongly relevant. The same is
true for the would be fixed line $t<1/2, \ \ \tilde\mu=0$, which is
present in the absence of the monopoles and describes the low
temperature massless phase of noncompact electrodynamics.

It is instructive to see how the RG equations formalize our
qualitative arguments of the previous section. In particular they make
clear the role of the point $t=1/2$ which in the absence of monopoles
would be the point where the charged induced term becomes relevant.
If one starts the evolution from the initial condition $t=1/2$,
$\tilde\mu\gg\tilde\zeta$, the running temperature will increase, and
$\tilde\mu$ will grow into the infrared, while $\tilde\zeta$ will grow
for a while until the running temperature reaches two. From this
point on $\tilde\zeta$ will decrease and the system will approach the
high temperature fixed point. This corresponds to the situation where
at the ``would be'' critical point $t=1/2$ the density of charged plasma
is 
so high that the mean distance between the particles is smaller than
the width of the confining string.

On the other hand if the initial
condition is $\tilde\mu<\tilde\zeta$, the temperature starts
decreasing, making the coupling $\tilde\mu$ immediately
irrelevant. Thus $\tilde\mu$ monotonically decreases to zero and
practically does not affect the flow of the other couplings which
steadily flow to the zero temperature fixed point. This is the
situation where at $t=1/2$ the density of charged plasma is low, which
is indeed true in our model. 

Interestingly enough, at $t=1$ the initial conditions in our model 
in the case of very heavy Higgs are
such that the RG flow starts almost exactly in the region of
attraction of the fixed point $T_{GG}$. At this value of temperature
the monopole and charge fugacities are equal. The only difference
between $\tilde \mu$ and $\tilde\zeta$ then is in the prefactors,
which are possibly different combinations of $M_W$ and $g^2$.
If $\tilde\mu$ and $\tilde\zeta$ were exactly equal, the
only thing that happens along the flow is that their values grow, but
$t$ does not change. Due to the small difference their values are
equal at a temperature which slightly differs from $1$
\be
{M_W\over T}={4\pi M_W\over g^2}+O(1), \ \ \ \ \ \ T={g^2\over
4\pi}[1+O({g^2\over M_W})]
\ee 
Thus the initial temperature for which the system is in the region of
attraction of $T_{GG}$ is slightly different. 
We conclude that the critical temperature of the Georgi-Glashow model
is indeed, up to corrections of order $g^2/M_W$, given by
$T_{GG}=\frac{g^2}{4\pi}$.

A comment is in order about our use of the perturbative
renormalization group equations. As written the equations (\ref{RG})
are only valid for small $\tilde\mu$ and $\tilde\zeta$. 
The initial values of these constants  in our model are indeed small,
and thus the flow is well described by these equations for quite a
while. The fixed points however 
correspond to large values of at least some parameters, and thus
quantitative details of the flow in this region of parameter space are
different. However the  existence of the three fixed points $T_0$,
$T_\infty$ and $T_{GG}$ is an exact statement. For $T_0$ and
$T_\infty$ this is a consequence of the fact that one of the
couplings vanishes ($\tilde\mu$ or $\tilde\zeta$) and thus these fixed
points are the same as in the corresponding sine-Gordon theory.
The existence of $T_{GG}$ is a consequence of the exact duality in
the model eq. (\ref{sinevortex}). The transformation
$2\chi\rightarrow\tilde\chi$
in the path integral leads to the same partition function but with new
parameters as follows $t\rightarrow 1/t$,
$\tilde\mu\rightarrow\tilde\zeta$,
$\tilde\zeta\rightarrow\tilde\mu$. This symmetry is
nothing but the famous $T$-duality (see for example \cite{Tduality}
and references therein). The point $T_{GG}$ is the fixed
point of this transformation and is thus guaranteed also to be a fixed
point of the exact renormalization group equations.

\subsection{The universality class of the phase transition}

Since the deconfining phase transition is due to the restoration of
$Z_2$ symmetry we expect that it is in the 2-dimensional Ising  
universality class.
This can be shown exactly by studying the theory eq. (\ref{sinevortex})
at the fixed point $t=1$.
The following discussion follows closely \cite{Nersesyan}[See also 
chapter 21 in \cite{Tsvelik}].

The theory described by eq. (\ref{sinevortex}) can be fermionized by using
the
standard bosonization/fermionization techniques. Since at $t=1$ both
cosine terms have dimension $1$, the resulting fermionic theory is a
theory of free massive fermions.
In our notations, the Dirac fermionic field is defined as
\be
\psi_{R}=a^{-1/2}i\exp[i(\chi+ {\tilde\chi\over 2})],
\hskip 1 cm \psi_{L}=a^{-1/2}\exp[- i(\chi- {\tilde\chi\over 2})],
.
\ee
The kinetic term in eq. (\ref{sinevortex}) then becomes the 
kinetic term of the field $\psi$, while the cosine terms become
\begin{eqnarray}
&&a^{-1}\cos 2\chi=i[\psi_R^\dagger\psi_L-H.c.]\nonumber\\
&&a^{-1}\cos \tilde\chi=i[\psi_R^\dagger\psi^\dagger_L-H.c.].
\end{eqnarray}
Thus the mass term in the fermionized Lagrangian is diagonalized
by introducing the real Majorana fermions
\be
\rho={\psi+\psi^\dagger\over \sqrt 2}, \ \ \
\sigma={\psi-\psi^\dagger\over i \sqrt 2}.
\ee
The mass of the fermion $\rho$ is $\mu a+\zeta a$, while
the mass of $\sigma$ is $\mu a-\zeta a$.
For $\mu=\zeta$ the model contains one massive and
one massless fermion. At the fixed point $\tilde\mu\rightarrow\infty$
the massive fermion decouples. Thus at the point $T_{GG}$ the theory
is that of one massless Majorana fermion. It is well know that this
theory precisely describes the critical point of the 2D Ising model.

Thus we indeed see that the phase transition is in the universality
class of the 2D Ising model.

\section{From the magnetic vortices to the Polyakov line}

It is standard practice to study the high temperature phase of
gauge field theories using the effective action in terms of the zeroth
component of the vector potential $A_0$. In this section we want to
relate our analysis which was performed for the phase of the vortex
operator $\chi$ to this standard procedure. 

The relation in
fact is not difficult to make.
The free energy of the charged $W^\pm$ bosons in
the Lagrangian eq. (\ref{sinevortex}) is given by the term
$\cos\tilde\chi$.
On the other hand this same free energy in the standard calculation is
represented by the insertion of the Polyakov line with charge two. We
thus identify the Polykov line with the exponential of the dual field
\be
P=\exp\{{i\over 2}\tilde\chi\}
\ee
and the dual field $\tilde\chi$ with the Abelian vector
potential
\be
\tilde\chi=2g\beta A_0
\label{iden}
\ee
Up to the monopole induced term, the
Lagrangian eq. (\ref{sinevortex}) is equivalent to the sine-Gordon
theory of the field $\tilde\chi$ in eq. (\ref{dualsG}). 
With the identification in eq. (\ref{iden}), this is
\be
{2\over T}(\partial_iA_0)^2+ \mu \cos ({2g\over T}A_0)
\ee
The monopole induced term also can be written in terms of the vector
potential. Its form in terms of the Polyakov line $P$ is precisely
the same as the form of the charge plasma induced term in terms of the
vortex field $V$. Recall that the origin of the plasma induced term
$\cos\chi$ in eq. (\ref{sinevortex}) is the dimensionally reduced
``Skyrme term'' of
eq. (\ref{skyrme}). 
Thus just like we derived the sine Gordon Lagrangian
eq.(\ref{sinevortex}) starting from the effective Lagrangian for the
vortex field, we can follow the same steps backwards but this time
expressing everything in terms of $P$.
Due to this duality between $V$ and $P$ we conclude that
the monopole induced term is the dimensionally reduced Skyrme term for
the Polyakov line. The full effective Lagrangian for the vector
potential is then
\be
L={2\over T}(\partial_iA_0)^2+\mu \cos ({2g\over T}A_0)+
{a^2\over 4\pi^2}\ln \left(\tilde\zeta \right)(\epsilon_{ij}
\partial_iP\partial_jP^*)^2
\label{a0}
\ee
The monopole in this Lagrangian is represented by a vortex of the
field ${g\over T}A_0$ with unit vorticity. 
The coefficient of the Skyrme term in eq. (\ref{a0}) is such that the
action of such a unit vortex is equal to the action of the ``core'' of
the monopole.

The cosine term in this expression is the potential for $A_0$ induced
by the nonvanishing density of the charged particles in the thermal
ensemble. Naturally, it contains the Debye ``electric'' mass term for
$A_0$ and also higher interactions.
Note that as opposed to strongly interacting theories, where similar
effective Lagrangians have been derived only in the derivative
expansion, the Lagrangian eq. (\ref{a0}) is valid on all distance scales
longer than $1/T$.
Thus in principle it can be used to calculate correlation
functions in a large momentum range. However, if one is interested in
the long distance behavior of the correlators, at temperatures above
$T_{GG}$ the higher derivative Skyrme term can be neglected. This is in
accordance with our analysis of the previous section which showed that
the monopole induced term is irrelevant above $T_{GG}$.

An interesting example of a correlation function is the correlator of
vortex operators $<V(x)V^*(y)>$. At low temperatures, since the
magnetic $Z_2$ symmetry is spontaneously broken this correlation function
tends to a constant at large separations. The corrections to this constant
value are given by the correlator of the field $\chi$. 
At zero temperature the correlator is
\be
<V(x)V^*(y)>={g^2\over 8\pi^2}\exp\{-{1\over 2}<\chi(x)\chi(y)>\}=
{g^2\over 8\pi^2}\exp\{-{16 \pi^2\over g^2}Y_3(x-y)\}
\label{t0}
\ee
where $Y_3(x-y)$ is the 3d Yukawa potential with the mass $M$.
At low temperature ($T<T_{GG}$) the infrared asymptotics 
(at distances $|x-y|>1/T$)
is instead
\be
<V(x)V^*(y)>={g^2\over 8\pi^2}\exp\{-{16 \pi^2T\over g^2}Y_2(x-y)\}
\label{lowt}
\ee
with $Y_2(x-y)$ - the 2D Yukawa potential with the mass which includes
the effects of integrating out the nonzero Matsubara modes, as
calculated in \cite{Zarembo}.
This expression follows from eq. (\ref{sinevortex}) neglecting the
$\cos\tilde\chi$ term, which is indeed negligible in the infrared.

At high temperatures $T>T_{GG}$ instead it is convenient to use
eq. (\ref{a0}) with the omission of the third term\footnote{Our
calculation is only valid in the range of 
temperatures below the Higgs expectation
value $h$. At $T>h$ the theory becomes essentially nonabelian, since
the Higgs particle as well as $W$ bosons are present in the ``vacuum''
in large numbers, and our methods become inapplicable.
In fact in this range there may be another phase
transition corresponding to the ``melting'' of the Higgs expectation
value. Our analysis in this and previous sections has nothing to say
about this regime.}. 
The calculation of
the correlator of the vortex operators proceeds along the lines of
\cite{kaks, Kovner1}. The insertion of $V(x)$ and $V^*(y)$ creates the
$Z_2$ domain wall stretching between the points $x$ and $y$. In terms
of the sine-Gordon theory eq. (\ref{a0}) this domain wall is just the
kink, and thus the domain wall tension is equal to the soliton mass $M_s$
\cite{Zamolodchikov}.
The correlator then is
\be
<V(x)V^*(y)>=\exp\{-M_s|x-y|\}
\label{ms}
\ee
with the soliton mass reading as
\be
M_s= a^{-1}{2\Gamma(p/2)\over \sqrt{\pi}\Gamma({{p+1\over 2}})}\Bigg[{ \pi
\Gamma({{1\over p+1}})\over  
\Gamma({{p\over p+1}})}\, 2\tilde\mu \Bigg]^{{p+1}\over 2},   
\ee
where
\be
p= { g^2\over 8\pi T - g^2}. 
\ee

Thus we find that the correlation function of
the vortex operators decreases exponentially in the high temperature
phase as it should in the phase with restored symmetry.
Note that were we to neglect the Debye mass term we would have found
that below $T_{BKT}$ the correlator tends to a constant at infinity,
while at $T>T_{BKT}$ it does decay at large distances but only as a
power, since the mass of the soliton would vanish.

We note that the result eq. (\ref{ms}) implies that in the deconfining
phase the monopoles are confined linearly. We remind the reader that
insertion of an operator $V$ is equivalent to creation of a
monopole\footnote{Strictly speaking $V$ creates a $Z_2$ monopole -
that is the monopole with half the magnetic charge than 
the 'tHooft-Polyakov solution. However the interaction between these
objects is qualitatively the same.}. 
Thus the logarithm of the correlation function can be
interpreted directly as the intermonopole potential. Indeed
at zero temperature one can read off the intermonopole potential from 
eq.(\ref{t0}). The potential is the Debye screened  3D Coulomb potential.
At nonzero temperature eq. (\ref{lowt}) gives it as the
2D Coulomb potential at short distances, again screened at distances 
$x^2>1/\zeta$. 
 At high temperature then we see from eq. (\ref{ms}) that the interaction
between the monopoles is linear with the ``string tension'' equal to
$M_s$. Thus the behaviour of the monopoles is in many senses ``dual''
to that of electric charges\footnote{Of course one always has to keep in
mind
that physically monopoles and charges are very different objects in
this models: the charges are particles, while the monopoles are
instantons.}.

Finally we wish to comment on the behaviour of the spatial Wilson loop
in this model. One expects the spatial Wilson loop to have an area law.
This area law has a simple interpretation in terms of the thermal 
ensemble of magnetic vortices \cite{reinhard}. The vortex picture is
generally true and as discussed in \cite{Kovner1}, holds even in 
theories where the Wilson loop
at zero temperature has a perimeter behaviour.
We have nothing to add to the discussion  of the vortex ensemble picture,
since our present approach is not formulated 
directly in these terms.
It is curious to note however, that the spatial string 
tension can not be calculated at all in the dimensionally
reduced theory of the field $\chi$, 
eq.(\ref{sinevortex}), although it can be easily shown that 
it does not vanish.
Consider the calculation of the Abelian Wilson loop $W(C)$. 
As discussed
in \cite{Kovner1}, it is the generator of the magnetic $Z_2$ 
transformation inside the contour $C$. The path integral
representation of this expectation value is
\begin{equation}
<W(C)>=\int d\chi 
\exp\Bigg\{-\int d^3x {g^2\over 8\pi^2} (\partial_\mu
\chi- j_\mu)(\partial_\mu\chi- j_\mu)+
+\cos 2\chi
\Bigg\}
\label{uexp}
\end{equation}
with
\begin{eqnarray}
j_\mu(x)&=&\pi\delta^{\mu 0}\delta(x_0),\ \ \ x\in S \\
&=& 0, \ \ \ x\notin S
\end{eqnarray}
and $S$ is the area bounded by $C$ lying in the $x_0=0$ plane.
This expression clearly can not be dimensionally reduced, since the 
external current $j_\mu$ has components at all Matzubara frequencies.
If we nevertheless naively disregard the higher Matzubara modes
of the $\chi$ field, we find\footnote{The contribution
of the zero Matzubara frequency is trivial, since the
path integral in this sector is dominated by the configuration
$\phi(x)=\pi, \ \ x\in S$ which has vanishing action.}
\be
W(C)=e^{-{g^2\over 8 a}S}
\ee
Since the cutoff on the reduced theory, $a$ is of
order $1/T$, parametrically this result is precisely what we expect
$\sigma\propto g^2T$. However to determine the exact numerical value
we would have to look into details of the nonzero Matsubara mode
configurations.

\section{Conclusions}

In this paper we have studied the deconfining phase transition in the
2+1 dimensional Georgi Glashow model. This model to this day remains
the only analytically solvable confining theory, and it is therefore
interesting to see how the deconfining transition occurs in this simple
yet nontrivial setting.

We found that the phase transition is indeed associated with the
restoration of the magnetic $Z_2$ symmetry with the universality class
of 2D Ising model. We have shown that the transition is driven by the 
appearance of the charged plasma rather than by the binding of the
monopoles. The presence of the monopoles is significant, since it
pushes the value of the critical temperature up by a factor of two (in
the infinitely heavy Higgs limit)
relative to the plasma transition in the noncompact electrodynamics.
We have derived the effective action in the high temperature phase
without recourse to the derivative expansion 
and have calculated the correlation function of the vortex
operators. These correlators decay exponentially above the phase
transition, which is consistent with the restoration of the magnetic
symmetry. This also shows that the monopoles are bound linearly at
high temperatures due the charged plasma effects, and these effects
are much more important than the monopole self interaction, which by
itself leads only to a logarithmic intermonopole potential.

We believe that these qualitative features are universal and should
generalize not only to the strongly interacting non Abelian theories
in 2+1 dimensions, but also in large measure to 3+1 dimensional QCD
along the lines discussed in \cite{Kovner1}.
Unfortunately we are not yet able to generalize the 
present quantitative analysis to these more interesting theories.

\leftline{\bf Acknowledgements}
We would like to thank Alesha Tsvelik for useful discussions. 
A.K. thanks Maxim Chernodub for useful discussions.
G.D. is supported  by the U.S. DOE grant DE-FG02-92ER40716.00, and thanks
PPARC for support through grant PPA/V/S/1998/00910 during a visit to
Oxford. A.K. is supported by PPARC. The research of  I.K. and B. T. are
supported by   PPARC Grant PPA/G/O/1998/00567.

\vskip 1cm




\myend
\begin{thebibliography}{99}


\bibitem{Polyakov}
A. M. Polyakov, Phys. Lett. {\bf B 59} (1975) 80;
Nucl. Phys. {\bf B 120} (1977) 429.

\bibitem{Zarembo}
N.O. Agasyan and K. Zarembo,
Phys.\ Rev.\  {\bf D57} (1998) 2475, hep-th/9708030.


\bibitem{'t Hooft}
G. 't Hooft,  Nucl. Phys. {\bf B 138} (1978) 1, Acta Phys. Austr. Suppl.
XXII (1980) 531.


\bibitem{Kovner}
A. Kovner and B. Rosenstein,  Int. J. Mod. Phys
{\bf A7} (1992) 7419.

\bibitem{Prasad}
M. K. Prasad and C. M. Sommerfield, Phys. Rev Lett. {\bf 35} (1975) 760;
E. B. Bogomolny, Sov. J. Nucl. Phys. {\bf 24} (1976) 861;
T. W. Kirkman and C. K. Zachos, Phys. Rev. {\bf D24} (1981) 999
 
\bibitem{greensite} J. Ambjorn and J. Greensite, JHEP {004} 9805
(1998); 

\bibitem{BKT}
V. L. Berezinskii, Sov. Phys. JETP {\bf 32} (1971) 493; J. M. Kosterlitz 
and D. J. Thouless, J. Phys. {\bf C7} (1973) 1181  


\bibitem{Kovner1}C.P. Korthals-Altes and A. Kovner,
Phys.Rev. {\bf D62} (2000) 096008. 

\bibitem{Kovner2} A. Kovner and B. Rosenstein, 
 JHEP {003} (1998) 9809.


\bibitem{Coleman}
S. Coleman, Phys. Rev. {\bf D11} (1975) 2088; S. Samuel,  Phys. Rev. 
{\bf D18} (1978) 1916; D. J. Amit, Y. Y. Goldschmidt and G. Grinstein, J.
Phys. {\bf A13} (1980) 585; D. Boyanovsky, J. Phys. {\bf A22} (1989)
2601.

\bibitem{Kadanoff}
J. V. Jos\'{e}, L. P. Kadanoff, S. Kirkpatrick and D. R. Nelson, 
Phys. Rev. {\bf B16} (1977) 1217; A. P. Young,  J. Phys. {\bf C11} (1978)
L4553; K. Huang and J. Polonyi,  Int. J. Mod. Phys
{\bf A6} (1991) 409.

\bibitem{Tduality}
A. Giveon, M. Porrati, E. Rabinovici, Phys.Rep. {\bf 244} (1994) 77.

\bibitem{Nersesyan} D.G. Shelton, A.A. Nersesyan and A.M. Tsvelik,
Phys. Rev. {\bf B53} 8521 (1996);


\bibitem{Tsvelik} A.O.Gogolin, A.A. Nersesyan and A.M. Tsvelik,
Bosonization and strongly correlated systems, Cambridge University
Press, 1998.

\bibitem{kaks} C.P. Korthals-Altes, A. Kovner and M. Stephanov, 
Phys. Lett. {\bf B469}, (1999) 205.

\bibitem{Zamolodchikov}
Al. B. Zamolodchikov,   Int. J. of Mod. Phys {\bf A 10} (1995) 1125

\bibitem{reinhard} M. Engelhardt, K. Langfeld, H. Reinhardt and O. Tennert
Phys.Rev.{\bf D61} (2000) 054504;
A. Hart, B. Lucini, Z. Schram and M. Teper;
JHEP {\bf 0006} (2000) 040;


\end{thebibliography}
